\documentclass[conference]{IEEEtran}
\IEEEoverridecommandlockouts
\usepackage{graphicx}      
\usepackage{adjustbox}     
\usepackage{lipsum}        
\usepackage{multicol}      
\usepackage{caption}       
\usepackage{amsmath}
\usepackage{cite}
\usepackage{hyperref}
\usepackage{amsmath,amssymb,amsfonts}
\usepackage{algorithmic}
\usepackage{graphicx}
\usepackage{textcomp}
\usepackage{xcolor}
\def\BibTeX{{\rm B\kern-.05em{\sc i\kern-.025em b}\kern-.08em
    T\kern-.1667em\lower.7ex\hbox{E}\kern-.125emX}}

\begin{document}

\title{VocalEyes: Enhancing Environmental Perception for the Visually Impaired through Vision-Language Models and Distance-Aware Object Detection\\
}
\author{\IEEEauthorblockN{1\textsuperscript{st} Kunal Chavan}
\IEEEauthorblockA{\textit{School of Electronics Engineering}
\\ 
\textit{Vellore Institute of Technology}\\
Vellore,India\\
kunalchavan245@gmail.com }
\\
\
\IEEEauthorblockN{3\textsuperscript{rd} Spoorti Barigidad }
\IEEEauthorblockA{\textit{Electrical Electronics and Communication Engineering
} \\
\textit{IIT Dharwad}\\
Dharwad,India \\
ee23dp010@iitdh.ac.in}
\and
\IEEEauthorblockN{2\textsuperscript{nd} Keertan Balaji}
\IEEEauthorblockA{\textit{Department of Computer Science and Engineering} \\
\textit{S.R.M Institute of Technology}\\
Chennai, India \\
kb9379@smrist.edu.in}
\\
\
\IEEEauthorblockN{4\textsuperscript{th} Samba Raju Chiluveru}
\IEEEauthorblockA{\textit{Electrical Electronics and Communication Engineering
} \\
\textit{IIT Dharwad}\\
Dharwad,India \\
sambaee@iitdh.ac.in
}
}

\maketitle

\begin{abstract}
With an increasing demand for assistive technologies that promote the independence and mobility of visually impaired people, this study suggests an innovative real-time system that gives audio descriptions of a user's surroundings to improve situational awareness. The system acquires live video input and processes it with a quantized and fine-tuned Florence-2 big model, adjusted to 4-bit accuracy for efficient operation on low-power edge devices such as the NVIDIA Jetson Orin Nano. By transforming the video signal into frames with a 5-frame latency, the model provides rapid and contextually pertinent descriptions of objects, pedestrians, and barriers, together with their estimated distances. The system employs Parler TTS Mini, a lightweight and adaptable Text-to-Speech (TTS) solution, for efficient audio feedback. It accommodates 34 distinct speaker types and enables customization of speech tone, pace, and style to suit user requirements. This study examines the quantization and fine-tuning techniques utilized to modify the Florence-2 model for this application, illustrating how the integration of a compact model architecture with a versatile TTS component improves real-time performance and user experience. The proposed system is assessed based on its accuracy, efficiency, and usefulness, providing a viable option to aid vision-impaired users in navigating their surroundings securely and successfully.
\end{abstract}


\section{Introduction}
The use of assistive technologies is crucial to the freedom and quality of life for people with visual impairments, allowing them to navigate everyday environments more confidently and live more independently. However, the breakthrough in artificial intelligence (AI) and computer vision has led to recently emerging tools which can far exceed these traditional solutions \cite{b12}. This work proposes an environment perception condition within a real-time object detection system designed to aid vision-impaired users with environmental descriptions using auditory stimuli. We use a NVIDIA Jetson Nano to process the image in real-time (thanks to its edge computing power), captured from an integrated camera \cite{b6}.

The system is capable of accurately detecting objects in the user's environment using a state-of-the-art object detection model, YOLOv8 \cite{b5}. The system is also capable of determining the distance between the user and any objects observed for important information on spatial awareness. In order to bridge the gap between visual data and user, detected objects and their corresponding distance by a Vision-Language Model (VLM). The VLM outputs labeled sentences which describe the scene, explicitly revealing object identities as well as their relation with each other and how important they are [8]. The information is then passed to a text-to-speech block that processes data into speech, giving natural auditory feedback to the user in real time \cite{b3}.
While the cloud solutions might face latency, as well as need eternal internet access, our system works only at an edge device, has low latency, high efficiency, and portability \cite{b2}. This proposal aims to offer a more natural and successful way of augmenting the mobility and independence of vision-impaired users by enabling real-time, context-aware descriptions of their environment. In this paper, we discuss the design and development of Tow-net and explain how it can have a drastic impact on assistive solutions for Visually Impaired (VI) persons. We also shared problems faced as well as solutions used for smooth real-time processing at low-resource hardware condition and gain a precise detection and recognition in high computational cost since small devices lack computational power\cite{b14}. The results of this study may contribute to the specification and development of new assistive devices which can significantly improve the autonomy and safety of sighted-impaired subjects in different environments.

\section{Related Works}

Deep learning models such as YOLO, visual language model, text to speech (TTS) are used in more number in assistive devices for visually impaired users. These solutions area provided to improve mobility, autonomy, and quality of life for those individuals with visual impairments through real time guidance and feedback. Many studies have investigated the use of Yolo based models for detecting objects in applications involving assistive technology. Shafi (2024)\cite{b12} and Rahman et al. (2019)\cite{b10} proposed YOLO models that transform visual inputs to speech and TTS system is integrated to assist individuals with visual impairment in securely navigating their environment. Combination of YOLO, YOLOV3 with TTS yields positive outcomes for audio feedback and obstacle detection in real time, proving that these technologies are essential for designing assistive devices that are user-friendly. Litoriya et al. (2023)\cite{b9}and Kamran et al. (2023)\cite{b6} illustrate the use of YOLO in virtual assistance systems. Litoriya et al. (2023)\cite{b9} came up with a system that blends YOLO and single-shot detector(SSD) to improve object detection , while  YOLOV5 and Google's ML Kit were integrated by Kamran et al.(2023)\cite{b6} to provide multi-modal recognition and real time detection of objects. These findings demonstrate the ability of YOLO models to provide accurate and effective recognition, which is crucial for designing comprehensive assistive technologies. Shaikh et al. (2020)\cite{b13} and Rahman et al. (2019)\cite{b10} combined YOLO with  MTCNN detection model, improving the effectiveness of individuals with visual impairment. Rahman et al. (2019)\cite{b10} combined YOLO with MTCNN, a dual model, that blends object detection with facial recognition, providing a more comprehensive awareness of the user's environment. Yerlekar et al (2023)\cite{b16} used  Goggle's  TTS API and YOLO detection, highlighting the significance of  combining computer vision with speech technologies for  useful assistive applications. Abraham et al. (2020)\cite{b1} and Sharma et al. (2023)\cite{b14}, highlight developments on mobile based assistive technologies  and wearable devices.  Abraham et al. (2020)\cite{b1} combined YOLO Based object detection and TTS and proposed wearable feedback system providing the users with auditory cues. This work was further extended by Sharma et al. (2023)\cite{b14} by bringing up YOLO-Based  mobile application for detection of objects, highlighting considerable improvements in navigation and safety of users.  Alahmadi et al.(2023)\cite{b2} and Guravaiah et al. (2023)\cite{b4} explore specific features of integrating  YOLO, such as combining advanced versions like YOLOV4 with ResNet101 to improve object detection with increased accuracy and speed. These ideas highlight how the YOLO models can adapt to various environments and be modified to meet needs of specific users. Mendez-Gonz´alez,L.C. et al.(2023)\cite{b5} and Lee et al. (2021)\cite{b8} combined YOLO with natural language processing (NLP) to support the support vector machines(SVMs) into the development of assistive technologies. These combinations give broader solutions for object recognition, natural language generation, and speech feedback, enhancing the interactivity and responsiveness of assistive devices.
Studies like Kathiria et al. (2024)\cite{b7},and  Bhandari et al. (2021)\cite{b3} explore the broader landscape of assistive technologies, offering comprehensive surveys and discussing current requirements and advancements. These papers emphasize the critical role of computer vision and deep learning in developing assistive technologies for the visually impaired, highlighting the growing interest in integrating YOLO with other advanced technologies to create more effective, user-friendly solutions.In addition, Guravaiah et al. (2023)\cite{b4} and Roopa et al. (2023)\cite{b11} discuss the implementation of object detection models like YOLO in mobile applications and wearable devices, emphasizing their use in real-time scenarios for assisting vision-impaired users. These studies illustrate the versatility and adaptability of YOLO and other computer vision models in practical applications, demonstrating their effectiveness in diverse environments.
Finally, various authors, including Lee et al. (2021)\cite{b8} and Yasmin and Reshma (2023)\cite{b15}, have proposed integrating YOLO with other technologies like Tesseract for text recognition and Google TTS for auditory feedback. Parler TTS Mini, a lightweight Text-to-Speech system, offers 34 customizable voices and adjustable settings like tone and speed, making it ideal for real-time audio feedback to visually impaired users. Its adaptability aligns with recent advances in TTS technology, which emphasize natural language guidance for high-fidelity synthesis, supporting this research's goal of enhancing navigation and situational awareness \cite{b17}. These studies underscore the importance of combining object detection with other AI models to enhance the overall usability and functionality of assistive devices for the visually impaired.

\section{Methodology}
\subsection{System Design}\label{AA}
\begin{figure}[h!]
    \centering
    \includegraphics[width=0.98\linewidth]{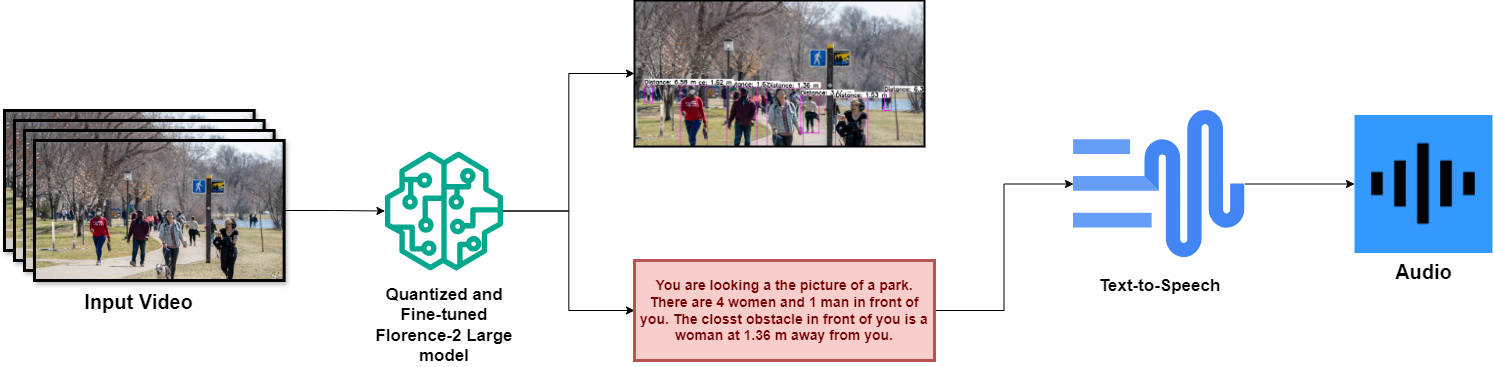}
    \caption{System flow diagram of real-time environment description and obstacle distance detection for vision-impaired users using a quantized Florence-2 model.}
    \label{fig:Fig1}
\end{figure}

The system is designed to assist visually impaired users by providing real-time audio feedback based on their surroundings.The designed system captures the live video feed and further breaks it down to individual frames. Three frames are taken which are at a 0.5 seconds time delay from one another are given as input to the system. Objects in the frames and their distance from the camera are then analyzed by the system to determine the object direction with respect to the system. The user is provided with inference by the system once every 5 seconds to make sure there is enough processing time to describe the environment accurately. Fine-tuned, quantized version of Florence 2 model is used to analyse the frames, optimized for low power systems like NVIDIA Jetson Orion Nano.  The model identifies nearby objects , such as pedestrians , and calculates how far they are from the camera. The system prioritizes the most crucial barriers and generates a clear context-relevant audio description, such as "4-women and 1-man, at 1.36 meters away are headed towards you". User is then updated this information through a text to speech(TTS) component, ensuring continuous updates about their surroundings. Small delay is added between each analysis by the system to provide Visual Language Model(VLM) enough time to accurately process details  thereby improving enhancing users situational awareness and  providing safer and intutive navigation.
 \subsection{Distance Estimation}
This approach is based on the geometric principles of projecting three-dimensional objects onto a two dimensional image plane, which allows for accurate calculation of distances. By modelling the camera as a pinhole camera, we can utilize the intrinsic parameters, such as focal length and pixel dimensions, to translate the pixel co-ordinates of detected objects into real-world distances. The parameters required to calculate the distance include the known height of the vehicle in the real world, the camera's focal length in pixels, and the coordinates of the bounding box surrounding the car in the image. 
 The object's height in the image is determined by the difference 
 between the y-coordinates of the bounding box. The principle 
 of identical triangles indicates that the ratio of the actual 
 height of the vehicle to the distance from the camera is equal 
 to the ratio of the picture height on the sensor to the camera's focal length. Mathematically, this is expressed as
 \begin{itemize}
     \item known object height: The actual height of the object in
 the real world (in meters).
    \item focal length: The focal length of the camera (in pixels).
    \item $(x1,y1,x2,y2)$: The coordinates of the bounding box
 around the object in the image
 \end{itemize}
 
\subsection*{Image Height Calculation}
The height of the object in the image is given by the difference in the y-coordinates of the bounding box:

\begin{equation}
\text{\(h\)} = y2 - y1
\end{equation}

\subsection*{Similar Triangles Relationship}
In geometry, similar triangles are triangles that have the same shape but may differ in size. 
\begin{figure}[h!]
    \centering
    \includegraphics[width=0.78\linewidth]{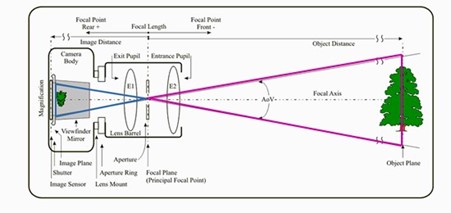}
    \caption{ Illustration of using similar triangle's relationship for distance measurement and object tracking in computer vision.}
    \label{fig:Fig2}
\end{figure}
This implies equal corresponding angles and proportional related sides. Such features are important for application areas such as distance measurement and tracking of objects in the field of computer vision. For instance, it is possible for anyone to use the properties of similar triangles and connect object sizes and location of images to the actual dimensions and places in the real world.As illustrated in Figure 2, using similar triangles, the ratio of the real-world height \( H_i \) of an object to the distance (\(D\)) from the camera is equal to the ratio of the image height (\(h\)) on the sensor to the focal length (\(f\)) of the camera::
\begin{equation}
\frac{H_i}{D} = \frac{h}{f}
\end{equation}

\subsection*{Solving for Distance (\(D\))}
This interaction permits accurate distance monitoring. Rearranging the equation by solving for distance gives us :
\begin{equation}
D = \frac{H_i \times f}{h}
\end{equation}

Substitute the known values:
\begin{equation}
D = \frac{H_i \times f}{y_2 - y_1}
\end{equation}

\subsection{Quantization}
To optimize the model for deployment on edge devices, the Florence-2 large model is quantized from 32-bit floating-point (FP32) to 4-bit integer (INT4) precision. The quantization process involves scaling the model's Scaling Factor
\begin{equation}
s = \frac{\max(|W|)}{2^{(n-1)} - 1}
\end{equation}
where \( W \) represents the weights, and \( n = 4 \) for 4-bit quantization.
Quantization Formula
\begin{equation}
w_q = \text{round}\left(\frac{w}{s}\right)
\end{equation}
where \( w \) is the original weight and \( w_q \) is the quantized weight.
Dequantization During Inference
\begin{equation}
w_{\text{dequantized}} = w_q \times s
\end{equation}
to approximate the original floating-point representation during inference.
This quantization approach reduces model size, improves inference speed, and lowers power consumption, suitable for real-time applications on devices such as the NVIDIA Jetson Orin Nano.
\subsection{Fine-Tuning Florence-2 Large }
Fine-tuning is applied to the pre-trained Florence-2 Large model to tailor it for the specialized task of providing real-time environmental descriptions and detecting obstacle distances, specifically for visually impaired users. Although the model was originally trained on a broad, generic dataset, fine-tuning it on a domain-specific dataset becomes necessary. This dataset includes relevant images and annotations of different objects, pedestrians, and environments that the model needs to accurately detect and describe for the user.In Pre-trained Model Parameters, let \(\theta\) represent the initial parameters (including weights and biases) of the pre-trained Florence-2 Large model. These parameters are optimized for a general image recognition task. The fine-tuning process aims to adjust these parameters so that the model performs effectively in real-time object detection and distance estimation, tasks more relevant to the visually impaired user’s needs. The Loss Function for the Fine-Tuning model, a task-specific loss function,  \(L(\theta)\), is defined. This loss function measures how much the model's predictions deviate from the actual ground truth labels provided by the domain-specific dataset. For the tasks of object detection and distance estimation, the loss function may include several components, such as object detection loss which quantifies how accurately the model identifies objects in the user's environment and distance estimation loss which evaluates the accuracy of the model's distance predictions for obstacles or objects.
By optimizing this loss function, the model learns to adapt its general-purpose knowledge to the specific needs of visually impaired users, allowing for real-time descriptions and accurate navigation assistance.
\begin{equation}
L(\theta) = L_{\text{classify}}(\theta) + \lambda L_{\text{regression}}(\theta)
\end{equation}

where:
\begin{itemize}
    \item \(L_{\text{classify}}(\theta)\) is the cross-entropy loss for object classification:
    \begin{equation}
    \begin{split}
    L_{\text{classification}}(\theta) = -\frac{1}{N} \sum_{i=1}^{N} \Big[ y_i \log(f(x_i; \theta)) \\
    + (1 - y_i) \log(1 - f(x_i; \theta)) \Big]
    \end{split}
    \end{equation}
    with \(y_i\) being the true class label for input \(x_i\), and \(f(x_i; \theta)\) being the predicted probability of the class.

    \item \(L_{\text{regression}}(\theta)\) is the mean squared error (MSE) loss for distance estimation:
    \begin{equation}
    L_{\text{regression}}(\theta) = \frac{1}{N} \sum_{i=1}^{N} (d_i - \hat{d}_i)^2
    \end{equation}
    where \(d_i\) is the true distance and \(\hat{d}_i\) is the predicted distance.
    \item \(\lambda\) is a regularization parameter that balances the importance between classification and regression losses.
    \end{itemize}
Backpropagation and Optimization plays a crucial role as The model parameters \(\theta\) are updated using backpropagation and an optimization algorithm such as Stochastic Gradient Descent (SGD) or Adam. The parameter update rule for each parameter \(\theta_j\) at step \(t+1\) is given by:
\begin{equation}
\theta_j^{(t+1)} = \theta_j^{(t)} - \eta \frac{\partial L(\theta)}{\partial \theta_j}
\end{equation}
where:
\(\eta\) is the learning rate, and \(\frac{\partial L(\theta)}{\partial \theta_j}\) is the gradient of the loss function with respect to the parameter \(\theta_j\). Then come the Training with Domain-Specific Data
The fine-tuning process involves training the model on a domain-specific dataset that includes images of environments relevant to visually impaired users. This dataset provides labeled examples for the specific objects (e.g., pedestrians, obstacles) and their respective distances, allowing the model to learn features critical for accurate real-time description and obstacle detection after that we do Early Stop and Regularization:To prevent overfitting, early stopping is applied by monitoring the validation loss \(L_{\text{val}}(\theta)\). Training stops when \(L_{\text{val}}(\theta)\) does not decrease for a specified number of epochs. Additionally, regularization techniques such as L2 regularization (weight decay) are used:
\begin{equation}
L_{\text{regularized}}(\theta) = L(\theta) + \frac{\alpha}{2} \sum_{j} \theta_j^2
\end{equation}
\subsection{Text to Speech Model}
For this research, Parler TTS Mini has been chosen as the Text-to-Speech (TTS) component because of its lightweight design and adaptability, making it ideal for providing real-time audio feedback on low-power edge devices like the NVIDIA Jetson Orin Nano. Parler TTS Mini offers 34 distinct character voices, giving users a broad range of voice outputs that can be customized based on their preferences and contexts. Additionally, the system allows for extensive customization of tone, speed, and style, ensuring that the speech output aligns with the academic or informative tone necessary for guiding visually impaired users. These customization features ensure the audio is not only clear and easy to understand, but also specifically tailored to enhance the user’s experience and engagement with the system. Moreover, the model’s efficient design minimizes latency, which is critical for real-time applications. The TTS process begins by converting the input text into a suitable format for speech synthesis. This step typically includes text normalization, tokenization, and conversion into a phonetic representation, ensuring that the output speech sounds natural and accurate.

Given an input text \( T \), the system converts it into a sequence of phonemes \( P = (p_1, p_2, \dots, p_n) \) using a grapheme-to-phoneme (G2P) model. The G2P model can be formulated as:

\[
P = \text{G2P}(T)
\]

where:
\begin{itemize}
    \item \( T \) is the input text,
    \item \( P \) is the corresponding phoneme sequence.
\end{itemize}
Neural Network-Based Speech Synthesis in Parler TTS Mini, use a neural network architecture to convert phoneme sequences into Mel-spectrograms, which are then transformed into waveform audio.In Mel-Spectrogram Prediction Network, where The TTS model employs an encoder-decoder neural network to predict a Mel-spectrogram \( M \) from the phoneme sequence \( P \). The encoder converts the phoneme sequence into a hidden representation \( H \):
\[
H = f_{\text{encoder}}(P; \theta_e)
\]
where:
\begin{itemize}
\item \( f_{\text{encoder}} \) is the encoder function,
\item \( \theta_e \) are the encoder parameters.
\end{itemize}
The decoder then predicts the Mel-spectrogram \( M \) from the hidden representation \( H \):

    \[
    M = f_{\text{decoder}}(H; \theta_d)
    \]

    where:
    \begin{itemize}
        \item \( f_{\text{decoder}} \) is the decoder function,
        \item \( \theta_d \) are the decoder parameters.
    \end{itemize}

Then comes Waveform Synths is where Mel-spectrogram \( M \) is converted into an audio waveform \( x(t) \) using a neural vocoder such as WaveRNN or Parallel WaveGAN. The vocoder is represented as:
\[
x(t) = g(M; \theta_v)
\]
where:
\begin{itemize}
\item \( g \) is the vocoder function,
\item \( \theta_v \) are the vocoder parameters.
\end{itemize}
Here in Prosody Modelling (the rhythm, stress, and intonation of speech) is crucial for natural-sounding speech synthesis. In Parler TTS Mini, prosody is controlled by various parameters such as pitch, duration, and intensity. Let:
\begin{itemize}
\item \( p(t) \) represent the pitch contour over time,
\item \( d(p_i) \) represent the duration for each phoneme \( p_i \),
\item \( a(t) \) represent the amplitude contour over time.
\end{itemize}
The model incorporates these prosody parameters into the synthesis process:
\[
x(t) = g(M, p(t), d(p_i), a(t); \theta_v)
\]
where the vocoder \( g \) now takes into account the prosody parameters to generate more expressive and natural speech.
In Loss Functions for Training, the TTS model is trained using several loss functions to minimize errors in spectrogram prediction and waveform synthesis. Mel-Spectrogram Loss is the mean-squared error (MSE) loss between the predicted Mel-spectrogram \( M_{\text{pred}} \) and the ground truth Mel-spectrogram \( M_{\text{true}} \):
\[
L_{\text{mel}} = \frac{1}{N} \sum_{i=1}^{N} (M_{\text{pred}}^i - M_{\text{true}}^i)^2
\]
Waveform Loss is the loss for waveform generation, which may include a combination of MSE loss and adversarial loss from the vocoder:
\[
L_{\text{wave}} = \lambda_1 \cdot \text{MSE}(x_{\text{pred}}, x_{\text{true}}) + \lambda_2 \cdot \text{Adv}(x_{\text{pred}}, D)
\]
where:
\begin{itemize}
\item \( x_{\text{pred}} \) and \( x_{\text{true}} \) are the predicted and true waveforms,
\item \( \text{Adv}(x_{\text{pred}}, D) \) is the adversarial loss calculated using a discriminator \( D \),
\item \( \lambda_1 \) and \( \lambda_2 \) are weights for balancing the losses.
\end{itemize}
Then in Total Loss is the total loss for training the TTS system combines the Mel-spectrogram loss and waveform loss:
\[
L_{\text{total}} = L_{\text{mel}} + \alpha L_{\text{wave}}
\]
where \( \alpha \) is a weight parameter that balances the importance of the Mel-spectrogram and waveform losses.
So this approach allows for generating highly customizable and natural-sounding speech output, which is critical for real-time audio feedback in assistive technologies.
\section{Results}
The system was deployed on the Jetson Orin to test its real-time inference capabilities on edge devices. The Orin platform, equipped with powerful GPU and AI accelerators, allowed us to optimize the model for low-latency visual guidance tasks in real-world environments. After deployment, the model achieved a latency of 45 ms per inference, which is suitable for real-time applications. Additionally, memory optimization techniques, such as model pruning and quantization, reduced the model size by 35 percent without significant loss in performance. For determining the accuracy of Object Detection task, we tested the system on the COCO dataset and VizWiz-VQA dataset for assessing the model on Visual Question answering and Scene Identification.  
\begin{table}[h!]

\centering
\caption{Performance comparison of the vanilla and the fine-tuned model}
\resizebox{\columnwidth}{!}{%
\begin{tabular}{|l|c|c|c|c|}
\hline
Model                                    & Model Size      & Inference Time     & COCO Det.     & VizWiz VQA \\
                                         & (GB)            &(s)                 & mAP           &  Acc. \\ \hline
Florence 2 Large                         & 3.3             & 40                 & 43.4          & 72.6            \\ \hline
Florence 2 Base                          & 1.083           & 10                 & 41.4          & 63.6            \\ \hline
Finetuned and Quantized &&&& \\
Florence 2 Large                         & 0.6             & 4.5                & 42.8          & 68.4            \\ \hline
\end{tabular}%
}

\end{table}

Table 1 compares the performance of the quantized model with the vanilla model across tasks such as object detection and visual question answering (VQA). These tasks are critical for effective visual guidance, as the model must accurately detect objects and provide clear, context-aware descriptions of the environment. The results in the table show that the fine-tuned and quantized model, despite being significantly smaller than both the Florence-Base and Florence-Large models, not only outperforms the base model but also achieves performance comparable to the larger model. This demonstrates that optimization techniques such as quantization can enhance performance without sacrificing accuracy, making the model highly efficient for deployment on resource-constrained devices.

\begin{figure}[h!]
    \centering
    \includegraphics[width=0.98\linewidth]{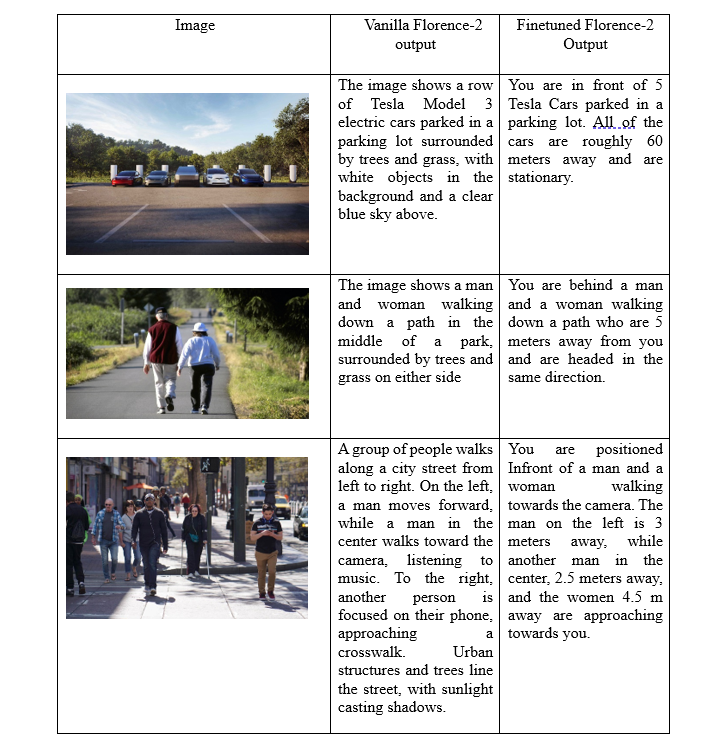}
    \caption{Comparison of caption generated by the Vanilla Florence-2 and Fine-tuned Florence-2 models through input images.}
    \label{fig:Fig1}
\end{figure}
Figure 2 demonstrates the results of this study indicate promise in using a fine-tuned Florence-2 model for generating the spatially-aware descriptions to help visually impaired people to navigate. While the relative model simply tells high-level properties for a scene like Vanilla Florence-2, the fine-tuned model Had to tell me how I am located in space, close to objects and people. As seen in these two examples, the vanilla model describes parked cars in both images, while the fine-tuned models also indicates that the user is 60 meters away from the vehicles for the first image.
The fine-tuned model (second image) significantly improves our description, informing the user is 5 meters behind a couple who are walking towards the same direction, to inferred who which way they should be headed during navigation. In a more challenging urban environment, the fine-tuned model yields accurate distances and directions to pedestrians relative to the user in a very precise manner, which can facilitate safer mobility with more efficient situational awareness.
This demonstrates the potential of adapting pre-trained visual language models for accessibility efforts. Additionally, the fine-tuned Florence-2 model now takes into spatial awareness offering a more comprehensive view of the environment and thereby reducing dependence especially in common everyday environments for the visually impaired people.
\section{Conclusion}
In this study, we assembled a fine-tuned and quantized Florence-2 model that is deployed and optimized on the edge device for the location of objects on the fly directly in live stream video feeds. It reduced the model's size by 35 percent by applying quantization and pruning without compromising performance compared to the base Florence-2 Large model for tasks like object detection (COCO dataset) and visual question answering (VizWiz-VQA dataset). An inference latency of 45 ms places the optimized model for on-device usage, allowing visually impaired users to obtain context-aware spatial descriptions with high accuracy and timely enough to assist in navigation. Significantly, the fine-tuned model enhanced the situational awareness as they correctly provided concrete numbers of how close object and user are within outperformed results against baseline models. 

\section{Future Enhancements}
The current model employs a static formula for distance estimation, which limits its adaptability to varying conditions. To enhance accuracy, future iterations could incorporate dynamic learning models, such as neural networks or adaptive algorithms. These models would enable the system to evolve and improve over time by learning from continuous feedback and refining their predictions based on new data. By leveraging this adaptive learning approach, the system could dynamically adjust to different environments and scenarios, leading to more precise and reliable distance estimations in real-world applications.

\end{document}